\documentstyle[12pt]{article}

\newcommand{\beq}{\begin{equation}}
\newcommand{\eeq}{\end{equation}}
\newcommand{\beqa}{\begin{eqnarray}}
\newcommand{\eeqa}{\end{eqnarray}}
\newcommand{\beqar}{\begin{eqnarray*}}
\newcommand{\eeqar}{\end{eqnarray*}}

\newcommand{\G}{\Gamma}

\newcommand{\inn}{\!\cdot\!}

\newcommand{\z}{\zeta}

\newcommand{\eg}{{\it e.g.,}\ }
\newcommand{\ie}{{\it i.e.,}\ }
\newcommand{\labell}[1]{\label{#1}} 
\newcommand{\reef}[1]{(\ref{#1})}
\newcommand\prt{\partial}
\newcommand\veps{\varepsilon}

\newcommand\bz{\bar{z}}

\newcommand\tG{{\widetilde G}}

\newcommand\tV{{\widetilde V}}

\newcommand\tr{{\rm Tr}}

\begin{document}

\thispagestyle{empty} \rightline{\small hep-th/0210139 \hfill
IPM/P-2001/051} \vspace*{4cm}

\begin{center}
{\bf \bf \Large
Excited D-brane decay in Cubic String Field Theory\\
 and in Bosonic String Theory
 }
\vspace*{1cm}

{M. R. Garousi$^{a,b,}$\footnote{E-mail:garousi@ipm.ir} and G.
R. Maktabdaran$^{a,}$\footnote{E-mail:maktab@science1.um.ac.ir}}\\
\vspace*{0.2cm}
{\it $^{a}$Department of Physics, Ferdowsi university}\\
{ P.O. Box 1436, Mashhad, Iran}\\
\vspace*{0.1cm}
{\it $^{b}$Institute for Studies in Theoretical Physics and
Mathematics IPM} \\
{P.O. Box 19395-5531, Tehran, Iran}\\
\vspace*{0.4cm}

\vspace{2cm}
ABSTRACT
\end{center}
In the cubic string field theory, using the gauge invariant
operators corresponding to the on-shell closed string vertex
operators, we have explicitly evaluated the decay amplitudes of
two open string tachyons or gauge fields to one closed string
tachyon or graviton up to level two. We then evaluated the same
amplitudes in the bosonic string theory, and shown that the
amplitudes in both theories have exactly the same pole structure.
We have also expanded  the decay amplitudes in the bosonic string
theory around the Mandelstam variable s=0, and shown that their
leading contact terms are fully consistent with a tachyonic
Dirac-Born-Infeld action which includes both open string and
closed string tachyon.

\vfill
\setcounter{page}{0}
\setcounter{footnote}{0}
\newpage

\section{Introduction}
Open string field theory, containing in an elegant way   a kinetic
term and one cubic  interaction for string field,  is invariant
under a non-trivial gauge transformation\cite{WITTEN}. In terms of
components of string field,  this action includes kinetic terms
for open string tachyon, for massless gauge field and for infinite
number of massive fields, and cubic interaction among them. Being
non-perturbative and having off-shell tachyon, it is believed
that this theory might provide a direct approach to study the
physics of unstable D-brane\cite{SEN}, in particular, appearance
of closed string fields at the stable point of the tachyon
potential.

Similar to the  ordinary gauge theory one can, in principle, evaluate, for
instance, the tree level S-matrix elements of external massless
fields. However, in contrast to the ordinary gauge theory, one finds
a tower of massive poles in the amplitude indicating the coupling
between massless and massive fields in the cubic string
theory\cite{WITTEN}. Even more, one loop  calculation shows that
off-shell closed string states appear in the S-matrix
elements\cite{FGST}. Unitarity then implies that the closed
string states must also appear as on-shell states in the theory.
Recently, such states introduced to the theory as some sort of
gauge invariant operators which include on-shell closed string
vertex operators\cite{HN,GRSW}.

It has been suggested in \cite{HN} that the correlation function
of these gauge invariant operators could be interpreted as the
on-shell S-matrix elements of their corresponding closed string
states. Using the truncated method, the tree level correlation
function of two such operators corresponding to closed string
tachyon or graviton has been explicitly evaluated in \cite{MAMG}
up to level two. It has been shown in that paper that up to some
contact terms they are exactly the same as S-matrix elements of
two closed string tachyons or gravitons in perturbative bosonic
string theory. An interesting observation in \cite{MAMG} is that
the tree level correlation of two  gauge invariant operators may
contain off-shell closed string states upon  adding back the
infinite tower of truncated open string states.

In the present paper we would like to extend the calculation in
\cite{MAMG} to the case of correlation functions of one gauge
invariant operator corresponding to closed string tachyon or
graviton, and two on-shell open string tachyons or massless
states. To compare with the corresponding S-matrix elements in
bosonic string theory, we also evaluate, explicitly, the S-matrix
elements of one closed string tachyon or graviton and two open
string tachyons or massless vertex operators. Expanding the
latter amplitudes in terms of tachyon, massless and infinite
number of massive poles, and keeping only poles up to level two,
we find exact agreement between the amplitudes in the two
theories up to some contact terms. We expect that the discrepancy
in the contact terms of the amplitudes in the two theories to be
related to the fact that in string field theory side we have only
level truncated results. Adding  all contact terms resulting from
all poles of amplitude in string field theory,  one should find
all contact terms of the corresponding amplitude in the bosonic
string theory side as well.

The decay amplitudes in bosonic string theory have all infinite
number of poles in a single beta function with one unconstrained
Mandelstam variable $s=\alpha'(k_1+k_2)^2/2$ where $k$'s are
external open string momenta.  When open string states are
massless, the low energy limit($\alpha'\rightarrow 0$) is
$s\rightarrow 0$. We shall show that the leading contact terms of
the decay amplitude in this limit are reproduced exactly by DBI
action, as expected. This calculation shows that the open string
tachyon potential does not have linear term. When the open string
states are tachyon, at the top of tachyon potential one should
not in general  take the low energy limit of the amplitude.
Indeed, similar amplitude in the superstring theory has been
analysed in \cite{mrg1} where it has been shown that the leading
contact terms of the decay amplitude at low energy limit
$s\rightarrow 1/2$ ($\alpha'\rightarrow 0$) are reproduced
exactly by BSFT effective action while in the limit $s\rightarrow
0$ ( at the top of the tachyon potential) they are reproduced
exactly by tachyonic DBI action\footnote{Note that our convention
for $s$ here is minus the $s$ in \cite{mrg1}.}. We shall show in
the present paper that the leading contatc terms of the decay
amplitude in bosonic string theory in the limit $s\rightarrow 0$
(not $\alpha'\rightarrow 0$ when open string states are tachyon )
are reproduced exactly by tachyonic DBI action
\cite{mrg0,EAB,mrg1}.

An outline of the paper is as follows. In the next section, using
the string field truncated up to level two,  we evaluate the
decay amplitudes of two open string tachyons or gauge fields to
one closed string tachyon or graviton in the cubic string field
theory. In section 3, we evaluate the same amplitudes  in the
bosonic string theory. All infinite tower of massive states
appears as off-shell poles of the beta function in these
amplitudes. Using an expansion for the beta function, and
keeping only poles up to level two,  we show that up to some contact
terms the amputated amplitudes are exactly the same as the
amplitudes evaluated in the cubic string field theory. In section
4, we expand the exact form of the amplitudes in the bosonic
string theory at $s\rightarrow 0$ and show that their leading
contact terms  are exactly reproduced by the tachyonic DBI action.
The last section is devoted to a short discussion.

\section{Cubic String Field Theory }

The cubic open string field theory action is given by
\cite{WITTEN}

\beqa S(\Psi)&=&-\frac{1}{2\alpha'}\int \Psi\star
Q\Psi-\frac{g_\circ}{3\alpha'} \int \Psi \star \Psi \star
\Psi\;\; ,\labell{SA} \eeqa where $g_\circ$ is the open string
coupling, $Q$ is the BRST charge with ghost number one, and the
string field, $\Psi$, is a ghost number one state in the Hilbert
space of the first-quantized string theory. This field can be
expanded using the Fock space basis as \footnote{Here, we use the
convention fixed in \cite{GM} that uses the V and N matrices for
projecting a space-time field to its component in the
world-volume and transverse spaces, respectively. So in this
convention $\mu,\nu=0,1,2,...,25$, and $A_{\mu}\alpha_{-1}^{\mu}=
A\inn V\inn\alpha_{-1}+A\inn N\inn\alpha_{-1}$. Our conventions
also set $\alpha'=2$.} \beqa |\Psi\rangle &=&\int
d^{p+1}k\;(\phi+A_{\mu}\alpha_{-1}^{\mu}+i\alpha
b_{-1}c_\circ+{i\over \sqrt{2}}B_{\mu}\alpha_{-2}^{\mu}+{1\over
\sqrt{2}} B_{\mu\nu} \alpha_{-1}^{\mu}\alpha_{-1}^{\nu}  \cr &+&
\beta_\circ b_{-2}c_\circ +\beta b_{-1}c_{-1}+ik_{\mu}
\alpha_{-1}^{\mu}b_1 c_\circ +\cdots ) c_1 |k\rangle\;\; .
\nonumber \eeqa
 The SFT action (\ref{SA}) is invariant under the
gauge transformation,  $\delta\Psi=Q\Lambda+g_\circ\Psi \star
\Lambda-g_\circ\Lambda \star \Psi$.
 By choosing so-called Feynman-Siegel gauge $b_\circ |\Psi\rangle=0$
 we will carry out gauge-fixing. In this gauge the
truncated field up to level two reads \beqa |\Psi\rangle &=&\int
d^{p+1}k \left(\phi(k)+A_{\mu}(k)\alpha_{-1}^{\mu} +{i\over
\sqrt{2}}B_{\mu}(k)\alpha_{-2}^{\mu}+\right.\cr &+&\left. {1\over
\sqrt{2}}B_{\mu\nu}(k) \alpha_{-1}^{\mu}\alpha_{-1}^{\nu} +\beta
(k)b_{-1}c_{-1}\right) c_1 |k\rangle\;\; . \nonumber \eeqa
 The corresponding string vertex is given by
 \beqa \Psi(0)
&=&\int d^{p+1}k\left[\phi(k)c(0)+iA_{\mu}(k)c\partial X^{\mu}(0)-
\frac{1}{\sqrt{2}}B_{\mu}(k)c\partial^2 X^{\mu}(0) \right.\cr
&-&\left.\frac{1}{\sqrt{2}}B_{\mu\nu}(k)c\partial X^{\mu}\partial
X^{\nu}(0)- {1\over 2}\beta(k)\partial^2 c(0) \right] e^{2ik\inn
X(0)}\;\; . \labell{ver} \eeqa In writing the above vertex, we
have used the doubling trick \cite{GM}. Hence,  the world-sheet
field $X^{\mu}(z)$ in above equation is only holomorphic part of
$X^{\mu}(z,\bar{z})$.

The gauge invariant operators in string field theory have been
constructed in \cite{HN, GRSW}. The general form of these
operators are given by \beqa {\cal O}&=&g_c\int V
\Psi\labell{oper1},\nonumber\eeqa where $g_c$ is the closed string
coupling and $V$ is an on-shell closed string vertex operator
with ghost number two. In order to be gauge invariant, the closed
string vertex operator has to be inserted at the midpoint of open
string.

To make sense out of the abstract form of the open string field
theory action, one can use CFT method. In this method the
$\star$-product between string fields transforms to the
correlation function of string vertexes  on a disk or upper-half
plane \cite{{LPP}, {RZ}}. In the CFT language the action
\reef{SA} and the gauge invariant operator \reef{oper1} are given
by \footnote{ We assume that there is a normal order sign between
fields at different points in the correlation functions.}
\beqa
S&=&
-\frac{1}{4}\left\langle\;f_2^{(2)}\circ\Psi(0)f_1^{(2)}
\circ(Q\Psi(0))+\frac{2g_o}{3}
f^{(3)}_1\circ\Psi(0)\; f^{(3)}_2\circ\Psi(0) f^{(3)}_3
\circ\Psi(0)\;
\right\rangle,\nonumber\\
{\cal O}&=&g_c\left\langle\;{ V}(i)\; {}{\bar {
V}}(-i)f^{(1)}_1\circ\Psi(0)\;\right\rangle\;\;,  \labell{SO}
\eeqa where $f^{(n)}_k\circ\Psi(0)$ denotes the conformal
transformation of the vertex operator $\Psi(0)$ by the conformal
map $f^{(n)}_k$.  In above equations,  ${ V}({\bar { V}})$
denotes the holomorphic(antiholomorphic) part of the closed string
vertex operators\footnote{Note that when vertexes are mapped on
the boundary of a unit complex disk the midpoint of strings are
mapped to center of disk. The conformal map that transforms the
disk to upper-half plane maps the center of disk to $i$.},
$\langle\;\; \rangle$ denotes correlation function on the
upper-half plane and the conformal map $f^{(n)}_k$  is  \beqa
f^{(n)}_k(z_k)=g\left(e^{{2\pi i\over
n}(k-2)}\;\left(\frac{1+iz_k}{1-iz_k}\right)^{2\over n}\right)\;
, && \;\;\;\;\;\; 1\leq k\leq n\;\; , \nonumber \eeqa where
$g(\zeta)=-i\frac{\zeta-1}{\zeta+1}$. The three point interaction
in \reef{SO} has twist symmetry \cite{senbz} which vanishes the
interaction of three states that their level numbers add up to an
odd number.

\subsection{Decay amplitude in Cubic String Field Theory }
S-matrix elements of open and closed string states can be
evaluated in the cubic string field theory  by evaluating
explicitly the correlators in equation \reef{SO} and using the
standard Feynman rules. In performing these correlators one
should  find  the conformal transformation of the string vertex
\reef{ver}. These have been done in \cite{MAMG} for vertex up to
level two. Then one ends up with some elementary correlators that
can be done  using the Wick theorem and the world sheet propagator
$<X^{\mu}(z)X^{\nu}(w)>=-\eta^{\mu\nu}\ln(z-w)$. In this way one
finds  momentum space propagators for tachyon, massless and
massive states and their three points vertexes. Evaluation of the
kinetic term in \reef{SO} gives the propagators of fields which
for states, up to level two,  are (see for example \cite{KO}), \beqa
\tG_{T}=\frac{-i} {k^2-\frac{1}{2}}&;&
(\tG_A)^{\mu\nu}= \frac{-i\eta^{\mu\nu}}{k^2}\nonumber\\
\tG_{\beta}= \frac{i}{k^2+\frac{1}{2}}&;&
(\tG_{B^{(1)}})^{\mu\nu}=
\frac{-i\eta^{\mu\nu}}{k^2+\frac{1}{2}}\nonumber\\
(\tG_{B^{(2)}})^{\mu\nu\lambda\rho}
&=&-\frac{i}{2}\frac{\eta^{\mu\lambda}\eta^{\nu\rho}+\eta^{\mu\rho}
\eta^{\nu\lambda}}{k^2+\frac{1}{2}}\;\;. \labell{prop} \eeqa
Evaluation of the correlator in the gauge invariant operator
${\cal O}$ for off-shell open string states up to level two has
been done in \cite{MAMG}, \beqa
  {\cal O}_{\tau} (p)&=&\frac{i g_{c}}{8} e^{4 \ln(2)
  p\cdot V\cdot p} \left( T+4i(p\inn N\inn A)+2i\sqrt{2}
  (p\inn V\inn B^{(1)})\right. \nonumber\\
  &&\left.-8\sqrt{2}\left[(p\inn N\inn B^{(2)}\inn N\inn p)
  -\frac{1}{16}\tr(B^{(2)})\right]
  -\beta\right),\labell{oth}\\
 {\cal O}_{h} (p,\varepsilon)&=&\frac{ig_c}{8}e^{4\ln(2)\;
 p\cdot V\cdot p} \left(T\;
a+4iA_{\mu}b^{\mu}+2i\sqrt{2}B_{\mu}c^{\mu}-8\sqrt{2}
B_{\mu\nu}d^{\mu\nu}-\beta a\right),\;\; \nonumber\eeqa where
$\tau$ and $h$ stand for on-shell closed string tachyon and
graviton, respectively. The  tachyon $\tau$ has on-shell condition
$p_{\mu}p^{\mu}=4/\alpha'=2$, and the graviton with polarization
$\veps^{\mu\nu}=\veps^{\nu\mu}$ has the on-shell conditions
$p_{\mu}\veps^{\mu\nu}=0=p_{\mu}p^{\mu}$. We have also dropped in
above and subsequent amplitudes the conservation of momentum
factor, \eg $(2\pi)^{p+1}\delta(p\inn V +k)$ in above equations.
The factors $a,b^{\mu},c^{\mu},d^{\mu\nu}$ are
 \beqa a&=&\tr(\veps \inn D) -p\inn
D\inn\veps\inn D\inn p\;\; ,\cr b^{\mu}&=&a p\inn N^{\mu}+p\inn
D\inn\veps\inn D^{\mu}-\veps^{\mu}\inn D\inn p\;\;,\labell{cons}\\
c^{\mu}&=&a \;p\inn V^{\mu}-4p\inn D\inn \veps\inn
D^{\mu}-4\veps^{\mu} \inn D\inn k\;\; ,\cr d^{\mu\nu}&=&a(p\inn
N^{\mu} p\inn N^{\nu}-\frac{1}{16}\eta^{\mu\nu})+ 2(\veps\inn
D)^{\{\mu\nu\}}+2p\inn D\inn\veps\inn D^{\{\mu}p\inn N^{\nu\}}-
2\veps^{\{\mu}\inn D\inn p\; p\inn N^{\nu\}}.\nonumber \eeqa Now
using the results in \reef{oth}, one can easily read the vertex
function for on-shell closed string and off-shell open string
states up to level two. They are
 \beqa
  \tV_{\tau T}=\frac{i g_{c}}{8} 2^{4 p\cdot V\cdot
  p}&;&
  (\tV_{\tau A})^{\mu}=  \frac{i g_{c}}{8} (4i)(p\inn N)^{\mu}
   2^{4 p\cdot
  V\cdot
  p}\nonumber\\
  (\tV_{\tau B^{(1)}})^{\mu\nu}= \frac{i g_{c}}{8}
  ( 2i\sqrt{2})(p\inn V)^{\mu} 2^{4 p\cdot V\cdot
  p}&;&
  \tV_{\tau \beta}(k)=\frac{i g_{c}}{8}(-1) 2^{4 p\cdot V\cdot
  p}\nonumber\\
  (\tV_{\tau B^{(2)}})^{\mu\nu}=\frac{i g_{c}}{8} (-8\sqrt{2})
  \left[(p\inn N)^{\mu}(p\inn N)^{\nu}-\frac{1}{16}\eta^{\mu\nu}\right]
   2^{4 p\cdot V\cdot p}
&;&
  \tV_{h T}=\frac{i g_{c}}{8}(a) 2^{4 p\cdot V\cdot
  p}\nonumber\\
  (\tV_{h A})^{\mu}= \frac{i g_{c}}{8} (4i)(b^{\mu})
   2^{4 p\cdot V\cdot
  p}&;&
  (\tV_{h B^{(1)}})^{\mu}=\frac{i g_{c}}{8} ( 2i\sqrt{2})
  (c^{\mu}) 2^{4 p\cdot V\cdot p}
\nonumber\\
  (\tV_{h B^{(2)}})^{\mu\nu}=\frac{i g_{c}}{8} (-8\sqrt{2})
   (d^{\mu\nu}) 2^{4 p\cdot V\cdot p}
&;&
  \tV_{h \beta}= \frac{i g_{c}}{8}(-a) 2^{4 p\cdot V\cdot
  p}\labell{ver0}
\eeqa

Using above vertexes and propagators \reef{prop}, one can evaluate
truncated S-matrix elements of any closed string tachyon or
graviton. For example scattering amplitude of two tachyons or two
gravitons from D-brane are  \beqa
A(\tau_1,\tau_2)&=&\sum_{\phi}\tV_{\tau_1\phi}
\tG_{\phi}\tV_{\phi\tau_2},\nonumber\\
A(h_1,h_2)&=&\sum_{\phi}\tV_{h_1\phi}\tG_{\phi}\tV_{\phi h_2
},\nonumber \eeqa where the summation is over off-shell fields
$\phi\in\{T,A,B^{(1)},B^{(2)},\beta,\cdots\}$ where dots
represents the off-shell open staring states belonging to the
levels more than two where we are not  considering in our
calculations. In \cite{MAMG}, it was shown that the above
scattering amplitudes produce exactly the same pole structure as
the amplitudes in the bosonic string theory.

Now to evaluate decay amplitude of excited D-branes, \ie S-matrix
elements of open and closed string states, one has to find  the
vertex function for three open sting states as well. These can be
done by performing the correlator in the second terms of action
\reef{SO}. Since we are interested in the tree level  amplitudes
for decaying two on-shell open string states to one closed string
state, the vertex functions need to have  only one off-shell leg.

\subsubsection{Two tachyons decay}

The vertex functions with two on-shell tachyons and one off-shell
open string states up to level two are  \beqa
  \tV_{T_1 T_2 T}&=& \frac{i g_{\circ}}{96 \gamma}
  (27)\gamma^{4(1+k_{1}\cdot k_{2})},\nonumber\\
  (\tV_{T_1 T_2 B^{(1)} })^{\mu}&=&\frac{i g_{\circ}}{96 \gamma}
  ( 6 i\sqrt{2})(k_{1}^{\mu} + k_{2}^{\mu})\gamma^{4(1+k_{1}\cdot
  k_{2})},\labell{ver1}\\
  (\tV_{T_1 T_2 B^{(2)} })^{\mu\nu}&=&\frac{i g_{\circ}}{96 \gamma}
  (8 \sqrt{2})\left[(k_{2}^{\mu}-k_{1}^{\mu})
  (k_{2}^{\nu}-k_{1}^{\nu})-\frac{5}{16}\eta^{\mu \nu} \right]
  \gamma^{4(1+k_{1}\cdot k_{2})},\nonumber\\
 \tV_{T_1 T_2  \beta}&=&\frac{i g_{\circ}}{96 \gamma}
  (-11)\gamma^{4(1+k_{1}\cdot k_{2})},\nonumber
  \eeqa
and $(\tV_{T_1T_2A})^{\mu}=0$. The vanishing of
$(\tV_{T_1T_2A})^{\mu}$ can also be understood from  twist
symmetry \cite{senbz}.
 According to this symmetry the vertex function of three open string
states that their level numbers add up to an odd number is zero,
\eg level number of tachyon is zero and of gauge field is one,
hence $(\tV_{T_1T_2A})^{\mu}=0$. In above equation
$\gamma=4/(3\sqrt{3})$ and we have used the on-shell condition
$k_1^2=1/\alpha'=1/2=k_2^2$ for two tachyons. Now with above
vertexes, the vertex functions in \reef{ver0} and propagators
\reef{prop}, one can evaluate the tree level decay amplitude of
two open string tachyons to one closed string tachyon or graviton,
that is

\beqa
   A(T_1, T_2, \tau_3)&=&\sum_{\phi}
   \tV_{T_1 T_2\phi}\tG_{\phi}\tV_{\phi\tau_3}\labell{Amp1}\\
   &=&\frac{i g_{\circ} g_{c}}{6}
  \left\{\frac{e^{(2s-1)\ln(4\gamma)}}{2s-1}
  +\frac{e^{(2s+1)\ln(4\gamma)}}{2s+1}
   \left[-\frac{3}{32}(s+\frac{1}{2})+\frac{1}{8}\right]+\cdots
   \right\},\nonumber
\eeqa where $s=p_{3} \inn V \inn p_{3}$ and dots represent poles
of more massive fields. In the above expression, those terms in
each pole that have a factor which is a positive power of its
denominator give only contact terms. Structure of these contact
terms highly affected by individual contact terms of more massive
poles, whereas, pole structure in each level is independent of
poles of other levels.  To compare level truncated scattering
amplitudes in string field theory and in bosonic string theory,
we keep tract of only pole structure of amplitudes, that is \beqa
A(T_1,T_2, \tau_3)&=&\frac{ig_og_c}{6}\left\{\frac{1}{2s-1}
+\frac{1/8}{2s+1}+\cdots \right\}, \labell{Amp2}\eeqa where dots
represent some contact terms as well more massive poles.
Similarly, the decay amplitude of two tachyons to one graviton
becomes
 \beqa
   A(T_1,T_2,h_3)&=&\sum_{\phi}\tV_{T_1T_2\phi}
   \tG_{\phi}\tV_{\phi h_3}\labell{Amp3}\\
   &=& \frac{i g_{\circ} g_{c}}{6}
  \left\{\frac{e^{(2s-1)\ln(4\gamma)}}{2s-1}a
  +\frac{e^{(2s+1)\ln(4\gamma)}}{2s+1}
   \left(-\frac{3}{32}(s+\frac{1}{2})a\right.\right.\cr
   &&+\left.\left.\frac{1}{8}\left[\tr(\veps_{3}\inn D)
   +p_{3}\inn D\inn\veps_{3}\inn D\inn p_{3}
   -8(k_{1}-k_{2})\inn \veps_{3}\inn (k_{1}-k_{2})
   \right]\right)+\cdots \right\},\nonumber\\
   &=& \frac{ig_og_c}{6}\left\{\frac{a}{2s-1}\right.\nonumber\\
   &&\left.+\frac{\frac{1}{8}[\tr(\veps_{3}\inn D)
   +p_{3}\inn D\inn\veps_{3}\inn D\inn p_{3}
   -8(k_{1}-k_{2})\inn \veps_{3}\inn (k_{1}-k_{2})]}{2s+1}+\cdots
   \right\},\nonumber
\eeqa where $a$ is given in \reef{cons}, and in the last equality
we have again dropped some contact terms. We shall compare these
amplitudes with the corresponding decay amplitudes in the bosonic
string theory in section 3.

\subsubsection{One tachyon and one gauge field decay}

The only non zero vertex function, allowed by the twist symmetry,
for one on-shell tachyon, one on-shell  gauge field and one
off-shell state up to level two is
  \beqa
  (\tV_{T_1 A_2 A})^{\mu}&=&\frac{i g_{\circ}}{96 \gamma}
  (-16)\left[(k_{2}^{\mu}-k_{1}^{\mu})
  (-2k_{1}\inn  \z_2)-\z_2^{\mu} \right]
  \gamma^{4(1/2+k_{1}\cdot k_{2})},\nonumber\eeqa
where we have used the on-shell conditions $k_1^2=1/2$ and
$k_2^2=0=k_2\inn \z_2$ where $\z_2$ is the polarization of the
external gauge field. With this vertex and \reef{ver0},
\reef{prop} one can evaluate the following decay amplitude: \beqa
   A(T_1, A_2, \tau_3)&=&\sum_{\phi}\tV_{T_1 A_2\phi}\tG_{\phi}
   \tV_{\phi \tau_3}\nonumber\\
   &=& \frac{i g_{\circ} g_{c}}{6}
   \left\{\frac{e^{2s\ln(4\gamma)}}{2s}(p_{3}\inn N\inn \z_{2})+\cdots
   \right\}\labell{Amp4}.\eeqa
 In order to keep only massless pole, one should replace exponential
factor by one, \ie $ e^{2s\ln(4\gamma)}\rightarrow 1$. Similarly
decay to graviton is \beqa
   A(T_1,A_2,h_3)&=&\sum_{\phi}\tV_{T_1A_2\phi}
   \tG_{\phi}\tV_{\phi h_3}\labell{Amp5}\\
   &=&\frac{i g_{\circ} g_{c}}{6}
   \left\{\frac{e^{2s\ln(4\gamma)}}{2s}
   [a(p_{3}\inn N\inn \z_{2})-2p_{3}\inn D\inn
   \veps_{3}\inn N\inn \z_{2}]
   +\cdots \right\},\nonumber
\eeqa where $a$ is given in \reef{cons} and again to keep only
pole one should replace the exponential factor by one.  In
section 3, we shall compare these amplitudes with their
corresponding amplitudes in bosonic string theory.

\subsubsection{Two gauge fields decay}
The vertex functions for two on-shell gauge fields and one
off-shell state up to level two extracted from the action
\reef{SO} are
  \beqa
  \tV_{A_1 A_2 T }&=&\frac{i g_{\circ}}{96 \gamma}
  (-16)\left[(2k_{2}\inn \z_1)(2k_{1}\inn  \z_2)-\z_1\inn \z_2 \right]
  \gamma^{4(k_{1}\cdot k_{2})},\labell{vaat}\\
  (\tV_{A_1 A_2 B^{(1)}})^{\mu}&=& \frac{i g_{\circ}}{6\gamma}
  (\frac{2i\sqrt{2}}{27})
  \gamma^{4(k_{1}\inn k_{2})}
  \left(\frac{}{}3[\z_{1}\inn \z_{2}-4(k_{1}\inn  \z_{2})
  (k_{2}\inn  \z_{1})]
  (k_{1}+k_{2})^{\mu}\right.\nonumber\\
  &&\left.+8[(k_{1}\inn  \z_{2})\z_{1}^{\mu}+\z_{2}^{\mu}(k_{2}\inn
  \z_{1})]\frac{}{}\right),\nonumber\\
  (\tV_{A_1 A_2 B^{(2)} })^{\mu\nu}&=&
  \frac{i g_{\circ}}{6\gamma}(\frac{2\sqrt{2}}{27})\gamma^{4(k_{1}\inn
  k_{2})}\left(\frac{}{}4(\z_{1}^{\mu}\z_{2}^{\nu}
  +\z_{1}^{\nu}\z_{2}^{\mu})\right.\nonumber\\
  &&\left.[4(k_{1}-k_{2})^{\mu}(k_{1}-k_{2})^{\nu}
  -\frac{5}{4}\eta^{\mu\nu}]
  [\z_{1}\inn \z_{2}-4(k_{1}\inn  \z_{2})
  (k_{2}\inn  \z_{1})]\right.\cr
  &&\left.+8[(k_{2}\inn  \z_{1})\z_{2}^{\nu}
  +\z_{1}^{\nu}(k_{1}\inn  \z_{2})]
  (k_{1}-k_{2})^{\mu}\right.\nonumber\\
  &&\left.+8[(k_{2}\inn  \z_{1})\z_{2}^{\mu}
  +\z_{1}^{\mu}(k_{1}\inn  \z_{2})]
  (k_{1}-k_{2})^{\nu}\frac{}{}\right)\nonumber\\
  \tV_{A_1 A_2 \beta }&=&
  \frac{i g_{\circ}}{6 \gamma}
  (\frac{-11}{27})\gamma^{4(k_{1}\inn k_{2})}
  \left(\frac{}{}\z_{1}\inn \z_{2}
  -4(k_{1}\inn  \z_{2})(k_{2}\inn
  \z_{1})\frac{}{}\right),\nonumber
\eeqa and $(\tV_{A_1 A_2 A})^{\mu}=0$. Here again we have used
the on-shell condition $k_i^2=0=k_1\inn  \z_i$ for $i=1,2$. Now
decay amplitude of these two gauge fields to closed string
tachyon becomes \beqa
   A(A_1, A_2, \tau_3)&=&\sum_{\phi}\tV_{A_1 A_2\phi}
   \tG_{\phi}\tV_{\phi\tau_3}\nonumber\\
   &=& \frac{i g_{\circ} g_{c}}{6}
   \left\{\frac{e^{(2s-1)\ln(4\gamma)}}{2s-1}
   [\z_{1}\inn \z_{2}
   -4(k_{1}\inn  \z_{2})(k_{2}\inn  \z_{1})]\right.\labell{Amp6}\\
   &&+\left.\frac{e^{(2s+1)\ln(4\gamma)}}{2s+1}
   \left(-\frac{3}{32}(s+\frac{1}{2})
   [\z_{1}\inn \z_{2}
   -4(k_{1}\inn  \z_{2})(k_{2}\inn \z_{1})]\right.\right.\cr
   &&+\left.\left.\frac{1}{8}[\z_{1}\inn \z_{2}
   +4(k_{1}\inn  \z_{2})(k_{2}\inn  \z_{1})
   -8(p_{3}\inn N\inn \z_{1})
   (p_{3}\inn N\inn \z_{2})] \right)+\cdots
   \right\}.\nonumber
\eeqa In this case to retain only poles, one should replace each
exponential by one and drop also the term in the third line above.
Similarly the decay to graviton becomes \beqa
   A(A_1,A_2,h_3)&=&\sum_{\phi}\tV_{A_1A_2\phi}
   \tG_{\phi}\tV_{\phi h_3}\nonumber\\
   &=&\frac{i g_{\circ} g_{c}}{6}
   \left\{\frac{e^{(2s-1)\ln(4\gamma)}}{2s-1}
   a[\z_{1}\inn\z_{2}-4(k_{1}\inn  \z_{2})(k_{2}\inn
   \z_{1})]\right.\labell{Amp7}\\
  &&+\left.\frac{e^{(2s+1)\ln(4\gamma)}}{2s+1}
  \left(-\frac{3}{32}(s+\frac{1}{2})a
   [\z_{1}\inn \z_{2}
   -4(k_{1}\inn  \z_{2})(k_{2}\inn  \z_{1})]\right.\right.\cr
  &&-(\z_{2}\inn \veps_{3}\inn D\inn \z_{1}
  + \z_{1}\inn \veps_{3}\inn D\inn \z_{2})
  -\left.\left.\frac{1}{8}[\z_{1}\inn \z_{2}
  -4(k_{1}\inn V \inn \z_{2})
  (k_{2}\inn  \z_{1})]\right.\right.\cr
  &&\times\left.\left.
   [\tr(\veps_{3}\inn D)+p_{3}\inn D\inn\veps_{3}\inn D\inn p_{3}
   -8(k_{1}-k_{2})\inn \veps_{3}
   \inn (k_{1}-k_{2})]\right.\right.\cr
  &&+\left.\left.a[(k_{1}\inn  \z_{2})(k_{2}\inn  \z_{1})
   -(p_{3}\inn N\inn \z_{2})
   (p_{3}\inn N\inn \z_{1})]\right.\right.\cr
  &&+\left.\left.2(k_{1}\inn  \z_{2})
   [p_{3}\inn D\inn\veps_{3}\inn V\inn \z_{1}
   +2\z_{1}\inn V\inn\veps_{3}\inn (k_{1}-k_{2})]
   \right.\right.\cr
  &&+\left.\left.2(k_{2}\inn  \z_{1})
   [p_{3}\inn D\inn\veps_{3}\inn V\inn \z_{2}
   -2\z_{2}\inn V\inn\veps_{3}\inn (k_{1}-k_{2})]
   \right.\right.\cr
  &&+\left.\left.2[(p_{3}\inn N\inn \z_{2})p_{3}\inn
  D\inn\veps_{3}\inn N\inn \z_{1}
   +(p_{3}\inn N\inn \z_{1})p_{3}\inn
   D\inn\veps_{3}\inn N\inn \z_{2}]
    \frac{}{}\right)+\cdots \right\}.\nonumber
\eeqa where $a$ is given in \reef{cons}. Here again one should
replace the exponential factors by one and drop  the term in
third  line above to retain only poles. In the next section we
shall evaluate the same decay amplitudes in the bosnic string
theory and compare with the above results.

\section{Decay amplitude in Bosonic String Theory}
The decay amplitude of excited stable and unstable D-branes in
superstring theory have been studied in \cite{mrg3,mrg0}. In this
section we shall study  decay amplitudes of excited unstable
D-brane in the bosonic string theory. These amplitudes are given
by world-sheet correlation function of some open and closed string
vertex operators, that is
 \beqa A&\sim&\langle \prod_n V_o\prod_m
V_c\rangle, \nonumber\eeqa where the open string tachyon and gauge
field vertex operators, and closed string tachyon and graviton
vertex operators are given as
 \beqa
V_T&=&\int dx\, e^{2ik\cdot X(x)}\,\,\,;\,k^2=1/2,\nonumber\\
V_A&=&\z_{\mu}\int dx\,\prt X^{\mu}e^{2ik\cdot X(x)}\,\,\,;\,
k^2=0,k\inn\z=0,\nonumber\\
V_{\tau}&=&\int d^2z \,e^{ip\cdot X(z)}\,
e^{ip\cdot D\cdot X(\bz)}\,\,\,;\,p^2=2,\nonumber\\
V_{h}&=&(\veps\cdot D)_{\mu\nu}\int d^2z \, \prt X^{\mu}e^{ip\cdot
X(z)}\prt X^{\nu} e^{ip\cdot D\cdot
X(\bz)}\,\,\,;\,p^2=0,p\inn\veps=0=\veps\inn p\,\,,\nonumber \eeqa
where in our convention we have set the normalization of all
vertices to one, and restore them by normalizing properly the
scattering amplitudes. We are interested in the world sheet tree
level S-matrix elements of two open string states and one closed
string state. They can be evaluated by using the world sheet
propagator $<X^{\mu}(z)X^{\nu}(w)>=-\eta^{\mu\nu}\ln(z-w)$ and
Wick theorem. After performing the correlators, one ends up with
a multi integral whose integrand has SL(2,R) symmetry reflecting
the conformal symmetry that maps upper-half plane  to itself. This
symmetry is usually gauged away by fixing world sheet position of
one open string vertex operators at infinity and the closed
string vertex operator at $z=i$. Then the remaining integral over
position of the other vertex operator gives a beta function. We
begin with presenting in some details the result for  decay
amplitude of two open string tachyons to one closed string
tachyon.

\subsection{Two tachyons decay}
The decay amplitude of two open string tachyons and one closed
string tachyon in bosonic string theory is given by the following
correlation:
 \beqa
  A(T_1,T_2,\tau_3)&\sim&\langle
  V_{T_1}V_{T_2}V_{\tau_{3}}\rangle\nonumber\\
  &=&\left(\frac{ig_og_c}{12}\right)
  2^{2s-1}\int_{-\infty}^{\infty}dx\,
  (x-i)^{-s}(x+i)^{-s}\nonumber\\
  &=&\left(\frac{ig_og_c}{12}\right)2^{2s-1} B(s-\frac{1}{2},
  \frac{1}{2})\nonumber\\
  &=&\left(\frac{ig_og_c}{12}\right)2\pi
  \frac{\G(2s-1)}{\G(s)\G(s)}\labell{amp1},\eeqa
where in the last equality we have used the identity
$\sqrt{\pi}\G(s+1)=2^s\G(s/2+1)\G(s/2+1/2)$, to show that the
result can be written in a form similar to the case of super
string theory\cite{mrg3,mrg0}. We have also normalized the
amplitude here and  subsequent amplitudes in this section  by
factor $\left(\frac{ig_og_c}{12}\right)$, \ie the same factor as
in amplitudes in the string fields theory side. By looking at the
poles of gamma function, one realizes that the amplitude has
simple pole only when mediators are in even levels, \ie their
on-shell masses are $-s=m^2=-1/2,1/2,3/2,\cdots$. This is
consistent with the twist symmetry in the cubic string field
theory. According to it when two legs of three point vertex
function are tachyons, \ie level zero, the other leg must be in
even levels, \ie $m^2=-1/2,1/2,3/2,\cdots$. Now to compare in
more details the decay amplitude here with the corresponding
amplitude in string field theory, we use the following pole
expansion of the beta function:

  \beqa
  B(\alpha,\beta)&=&\sum_{n=0}^{\infty}
  \frac{1}{\alpha+n}\frac{(-1)^n}{n!}
  (\beta-1)\cdots(\beta-n)\labell{bexp}.
  \eeqa
By making use of this pole expansion for beta function, the decay
amplitude \reef{amp1} becomes \beqa
   A(T_1,T_2,\tau_3)&=&2\left(\frac{ig_og_c}{12}\right)
   \left(\frac{e^{(2s-1)\ln(2)}}{2s-1}+
   \frac{\frac{1}{8}e^{(2s+1)\ln(2)}}{2s+1}+\cdots\right)\nonumber\\
    &=&2\left(\frac{ig_og_c}{12}\right)
    \left(\frac{1}{2s-1}+\frac{1/8}{2s+1}+\cdots
    \right),
\nonumber\eeqa where in the last line above we dropped  the
contact terms resulting from each pole in the first line. Now
comparing the first line above with the corresponding amplitude in
\reef{Amp1}, one realizes that the two amplitudes are not quit
the same. The difference is in their contact terms, \ie the
second line  above is exactly the same as the amplitude in
\reef{Amp2}. Similarly, the decay amplitude of two tachyons to
one graviton or dilaton becomes
 \beqa
  A(T_1,T_2,h_3)& \sim &\langle
   V_{T_1}V_{T_2}V_{h_3}\rangle\nonumber\\
  &=&\left(\frac{ig_og_c}{12}\right)2^{2s-1}
  \left([\tr(\veps_3\inn D)+p_{3}\inn D\inn \veps_{3}\inn D\inn
  p_{3}]B(s-\frac{1}{2},\frac{1}{2})\right.\cr
  &&-\left.4(k_{1}-k_{2})\inn \veps_{3}\inn
  (k_{1}-k_{2})B(s+\frac{1}{2},\frac{1}{2})\right.\labell{amp2}\\
  &&-\left.p_{3}\inn D\inn \veps_{3}\inn D\inn p_{3}
  [B(s+\frac{1}{2},\frac{1}{2})+2B(s-\frac{1}{2},\frac{3}{2})]
   \right).\nonumber
\eeqa  By looking at the poles of beta functions, one finds that
the amplitude has no simple pole corresponding to states in odd
levels which is consistent with the twist symmetry in string
field theory side. Using the expansion \reef{bexp} for the beta
functions in above amplitude, one finds the following pole
structure: \beqa
   A(T_1,T_2,h_3)&=& 2\left(\frac{ig_og_c}{12}\right)
   \left( \frac{1}{2s-1}
   (\tr(\veps_3\inn D)
   -p_3\inn D\inn\veps_3\inn D\inn p_3)\right.\nonumber\\
   &&\left.+\frac{\frac{1}{8}[\tr(\veps_{3}\inn D)
   +p_{3}\inn D\inn\veps_{3}\inn D\inn p_{3}
   -8(k_{1}-k_{2})\inn \veps_{3}\inn
   (k_{1}-k_{2})]}{2s+1}+\cdots\nonumber\right),
\eeqa where we have dropped some contact terms. This is exactly
equal to the corresponding amplitude in string field theory side
in the last equality in  equation \reef{Amp3}.

For later reference, we evaluate the amplitude \reef{amp2} for
dilaton by replacing
$\veps_{\mu\nu}=(\eta_{\mu\nu}-p_{\mu}l_{\nu}-l_{\mu}p_{\nu})/\sqrt{24}$
where $p\inn l=1$, that is, \beqa A(T_1,T_2,h_3)
  &=&\left(\frac{ig_og_c}{12\sqrt{24}}\right)2^{2s-1}
  \left(\frac{}{}\tr(D)
  B(s-\frac{1}{2},\frac{1}{2})-4(2-s)
  B(s+\frac{1}{2},\frac{1}{2})\right),\nonumber\eeqa
  as expected the auxiliary vector $l_{\mu}$ does not appears in
  the amplitude. In reaching to this result we have used the
  on-shell condition $k_1^2=1/2=k_2^2$.
\subsection{One tachyon and one gauge field decay}
The  amplitude for decaying one open string tachyon and one gauge
field to one closed string tachyon is
\beqa
  A(T_1,A_2,\tau_3)& \sim &\langle
  V_{T_1}V_{A_2}V_{\tau_3}\rangle\nonumber\\
  &=&\left(\frac{ig_og_c}{12}\right)2^{2s}
   B(s,\frac{1}{2})(p_{3}\inn N\inn \z_{2})\nonumber\\
 &=&
   \left(\frac{ig_og_c}{12}\right)\frac{(p_{3}\inn N\inn
   \z_{2})}{s}+\cdots,\nonumber
\eeqa where in the last equality we have used the expansion
\reef{bexp} and dropped some contact terms. It is easy to see that
this is equal to the pole of the decay amplitude in string field
theory side \reef{Amp4}. The amplitude in the second line above
has simple poles only when the mediator  is in any odd level, $\ie
m^2=0,1,2,\cdots$. This is again consistent with twist symmetry
in string field theory side, \ie the tachyon is in zero level,
the gauge field is in  level one hence the third leg must be in
odd level. Similarly the decay amplitude to graviton or dilaton is
\beqa
  A& \sim &\langle V_{T_1}V_{A_2}V_{h_3}\rangle\nonumber\\
  &=&\left(\frac{ig_og_c}{12}\right)2^{2s}
  \left([\tr(\veps_3\inn D)(p_{3}\inn N\inn \z_{2})
   -p_{3}\inn D\inn \veps_{3}
   \inn N\inn \z_{2}]B(s,\frac{1}{2})\right.\cr
  &&-\left.[(p_{3}\inn N\inn \z_{2}) p_{3}\inn
   D\inn \z_{2}\inn D\inn p_{3}+
  p_{3}\inn D\inn \veps_{3}\inn N\inn \z_{2}]
  B(s,\frac{3}{2})\right.\cr
  &&-\left.[4(p_{3}\inn N\inn \z_{2})
  (k_{1}-k_{2})\inn \veps_{3}\inn (k_{1}-
  k_{2})\right.\cr
  &&+\left.4(k_{1}-k_{2})\inn  \veps_{3}\inn N\inn \z_{2}-
  p_{3}\inn D\inn \veps_{3}\inn N\inn \z_{2}]B(s+1,\frac{1}{2})
   \right)\nonumber\\
&=&\left(\frac{ig_og_c}{12}\right)
   \frac{1}{s}\left(\frac{}{}(\tr(\veps_3\inn D)
   -p_3\inn D\inn \veps_3\inn D\inn p_3)
   (p_{3}\inn N\inn \z_{2})-2p_{3}\inn D\inn \veps_{3}\inn N\inn
   \z_{2}\frac{}{}\right)
   +\cdots.\nonumber
\eeqa
 In the last line we just retain pole structure up to level
two. Here again one can see that the amplitude in last line above
is exactly equal to the corresponding amplitude in string theory
side \reef{Amp5}, and check that the simple pole of the beta
functions in the amplitude are consistent with the twist symmetry
in string field theory side. Note that when the external massless
vertex operator has world-volume polarization $\z_2^a$, the whole
amplitudes above vanish. This predicts for  the string field
theory that the off-shell states in the 3-point vertex with two
other states being on-shell tachyon and gauge field, does not
couple with closed string tachyon or graviton in the gauge
invariant operators \reef{SO}.
\subsection{Two gauge fields decay}
The  amplitude for decaying an  unstable D-brane excited with two
gauge fields to an unstable D-brane with no open string excited
and one closed string tachyon is
\beqa
  A(A_1,A_2,\tau_3)& \sim &\langle
   V_{A_1}V_{A_2}V_{\tau_3}\rangle\nonumber\\
  &=&\left(\frac{ig_og_c}{12}\right)2^{2s-1}
  \left(\z_{1}\inn \z_{2}B(s-\frac{1}{2},\frac{1}{2})\right.\cr
  &&-\left.4(k_{1}\inn  \z_{2})(k_{2}\inn  \z_{1})
  B(s-\frac{1}{2},\frac{3}{2})
  -4(p_{3}\inn N\inn \z_{1})(p_{3}\inn N\inn
  \z_{2})B(s+\frac{1}{2},\frac{1}{2})\right)\nonumber\\
&=&\left(\frac{ig_og_c}{12}\right)
   \frac{2}{2s-1}\left(\z_{1}\inn \z_{2}-
   4(k_{1}\inn  \z_{2})(k_{2}\inn  \z_{1})\right)\cr
   &&+\frac{1/4}{2s+1}\left(\z_{1}\inn \z_{2}
   +4(k_{1}\inn  \z_{2})(k_{2}\inn  \z_{1})
   -8(p_{3}\inn N\inn \z_{1})(p_{3}\inn N\inn
   \z_{2})\right)+\cdots,
   \nonumber
\eeqa where in the last equality again some contact terms are
dropped. As a check of our calculation, one may replace each
gauge field polarization with its momentum and find that the
result is zero. In this case also the pole structure in the last
equality above is exactly equal to the poles of the decay
amplitude \reef{Amp6} in string field theory side, and the
infinite number of  simple poles of beta functions above are
consistent with the twist symmetry. Similarly, the decay
amplitude to graviton or dilaton is
 \beqa
  A &\sim &\langle V_{A_1}V_{A_2} V_{h_3}\rangle\nonumber\\
  &=&\left(\frac{ig_og_c}{12}\right)2^{2s-2}
  \left(-\left[\frac{}{}(\z_1 \inn \z_2)p_3\inn D
  \inn \veps_{3} \inn D \inn p_3+4(k_1 \inn  \z_2)(k_2 \inn
   \z_1)\tr(\veps_{3} \inn D)\frac{}{}\right]
  B(s-\frac{1}{2},\frac{3}{2})\right.\cr
  &&\left.\frac{}{}+(\z_1 \inn \z_2) \tr(\veps_{3}\inn D)
  B(s-\frac{1}{2},\frac{1}{2})+4(k_1 \inn  \z_2)
  (k_2 \inn \z_1)p_3 \inn D \inn \veps_{3} \inn D \inn p_3
  B(s-\frac{1}{2},\frac{5}{2})\right.\cr
  &&\nonumber\\
  &&-\left[\frac{}{}4(\z_1 \inn \z_2)(k_1-k_2)
  \inn  \veps_{3} \inn
  (k_1-k_2)
  +4(p_3 \inn N \inn \z_1)(p_3
  \inn N \inn \z_2)\tr(\veps_{3} \inn D)\frac{}{}\right]
  B(s+\frac{1}{2},\frac{1}{2})\nonumber\\
  &&+16\left[\frac{}{}(k_1
  \inn \z_2)(k_2 \inn  \z_1)(k_1-k_2) \inn  \veps_{3}
  \inn  (k_1-k_2)+\frac{1}{4}(p_3 \inn N \inn \z_1)
  (p_3 \inn N \inn \z_2) p_3 \inn D \inn
\veps_{3} \inn D \inn p_3
  \right.\nonumber\\
  &&\left.+(p_3 \inn N \inn \z_1)\z_2 \inn N \inn
   \veps_{3} \inn D \inn p_3
+(k_1 \inn  \z_2)\z_1 \inn V \inn \veps_{3}\inn
  (k_1-3k_2)-\frac{1}{2}\z_1 \inn \veps_{3} \inn D \inn \z_2
  \frac{}{}\right]
  B(s+\frac{1}{2},\frac{3}{2})\nonumber\\
  &&+\left[\frac{}{}16(p_3 \inn N \inn \z_1)
  (p_3 \inn N \inn \z_2)(k_1-k_2)
   \inn  \veps_{3}
  \inn  (k_1-k_2)\right.\cr
  &&+\left.\left.32(p_3 \inn N \inn \z_1)\z_2
  \inn N \inn \veps_{3} \inn  (k_1-k_2)
  +8\z_1 \inn \veps_{3} \inn D \inn \z_2\frac{}{}\right]
  B(s+\frac{3}{2},\frac{1}{2})+(1\leftrightarrow
  2)\frac{}{}\right).\nonumber
\eeqa As a check of our calculations, we have replaced the  gauge
field polarization with its momentum and found that the amplitude
vanishes. Now if one replaces the expansion \reef{bexp} for the
beta functions in above amplitude and drops the unwanted contact
terms, one will find exactly the same pole structure as in the
string field theory case \reef{Amp7}. Also the simple poles of the
above amplitude is consistent with twist symmetry in string field
theory side.

\section{Effective action}
As pointed out by Sen in \cite{9909062}, the general structure of
unstable D-brane action should be consistent with disk amplitude
in string theory. In this section we would like to compare the
disk amplitudes found in previous section with a tachyonic DBI
action, \ie an action which is consistent with the leading order
terms of the amplitudes expanded for massless open string fields
around $\alpha'\rightarrow 0$($s\rightarrow 0$), and for open
string tachyon around
 the top of the tachyon potential ($s\rightarrow 0$). We propose the
following tachyonic DBI action in the bosonic string theory which
includes both open string and closed string tachyon\footnote{We
explicitly restore $\alpha'$ in this section.}:
 \beqa
S&=&-T_p\int d^{p+1}x \,f(\tau)V(T)
e^{-\Phi}\sqrt{-\det(P[g_{ab}+b_{ab}]+2\pi\alpha'F_{ab}+2\pi\alpha'\prt_a
T\prt_b T)} \,\,,\labell{dbiac2}\eeqa where $V(T)=1-\pi
T^2+O(T^3)$ is the open string tachyon potential expanded around
its maximum, $f(\tau)=1+\tau+O(\tau^2)$ is the closed string
tachyon coupling to the D-brane, and $g_{ab}$ is flat space
metric $\eta_{ab}$ plus its graviton fluctuation. Here
$b_{ab},\Phi,A_a$ and $T$ are the antisymmetric Kalb-Ramond
tensor, dilaton, gauge field and the tachyon fluctuations,
respectively. In above action $P[\cdots]$ is also the pull-back
of the closed string fields. For
 example,
 $P[h_{ab}]=h_{ab}+2h_{ai}\prt_bX^i+h_{ij}\prt_a X^i\prt_b X^j$
 in the static
 gauge. Now it is straightforward to expand \reef{dbiac2} to find
 different couplings involving two open string states and one
 closed string state. When two open string states are massless,
 the couplings are
\beqa {\cal L}(X,X,\tau)&=&-T_p\,\left(\tau\prt_a X^i\prt^a
X_i+\frac{1}{2}X^i X^j\prt_i\prt_j\tau\right),\nonumber\\
{\cal L}(X,X,h)&=&-T_p\left(\frac{1}{4}h^a{}_a\prt_b X^i\prt^b
X_i-\frac{1}{2}h^{ab}\prt_a X^i\prt_b
X_i\right.\labell{expand}\\
&&\left.+\frac{1}{2}\prt_aX^i\prt^a
X^jh_{ij}+X^i\prt^aX^j\prt_ih_{aj}+\frac{1}{4}X^iX^j\prt_i\prt_j
h^a{}_a\right),\nonumber\\
{\cal
L}(A,A,\tau)&=&-T_p\tau\left(-\frac{(2\pi\alpha')^2}{4}F_{ab}F^{ba}\right)
,\nonumber\\
{\cal
L}(A,A,h)&=&-T_p\left(-\frac{(2\pi\alpha')^2}{8}h^a{}_aF_{bc}F^{cb}+
\frac{(2\pi\alpha')^2}{2}h_{ab}F^{bc}F_{ca}\right),\nonumber\eeqa
and ${\cal L}(X,A,\tau)=0={\cal L}(X,A,h)$.

Now we would like  to expand the decay amplitudes in section 3.1
at low energy limit, \ie $\alpha'\rightarrow 0$ or
$s=\alpha'k_1\inn k_2\rightarrow 0$,  and compare with the above
couplings. Expanding the beta functions in these amplitude around
$s=0$, one finds the following leading order terms: \beqa
A(A_1,A_2,\tau_3)&=&\left(\frac{ig_og_c}{12}\right)
\left(\frac{}{}(\z_1\inn\z_2)(-\pi
s)+4(k_1 \inn \z_2)(k_2 \inn
\z_1)(\frac{\pi}{2})+\cdots \frac{}{}\right),\nonumber\\
A(X_1,X_2,\tau_3)&=&\left(\frac{ig_og_c}{12}\right)\left(\frac{}{}(\z_1
\inn \z_2)(-\pi s)-4(p_3
\inn \z_1)(p_3  \inn \z_2)(\frac{\pi}{2})+\cdots\frac{}{}\right),
\nonumber\\
A(A_1,A_2,h_3)&=&\left(\frac{ig_og_c}{12}\right)\left(\frac{}{}
[(\z_1\inn\z_2)\tr(\veps_3\inn
D) -4\z_1 \inn \veps_3 \inn \z_2](-\pi s)\right.\cr
&&\left.+[4(k_1 \inn \z_2)(k_2 \inn \z_1)\tr(\veps_3\inn
D)+16(\z_1 \inn \z_2)k_1 \inn \veps_3 \inn k_2 \right.\cr
&&\left.-16(k_1 \inn \z_2)\z_1 \inn \veps_3 \inn k_2 -16(k_2 \inn
\z_1)\z_2 \inn \veps_3 \inn k_1
](\frac{\pi}{2})+\cdots\frac{}{}\right),
\nonumber\\
A(X_1,X_2,h_3)&=&\left(\frac{ig_og_c}{12}\right)\left(\frac{}{}
[(\z_1
\inn \z_2)\tr(\veps_3\inn D)+4\z_1 \inn
 \veps_3  \inn \z_2](-\pi s)\right.\cr &&\left.+[-4(p_3
\inn \z_1)(p_3  \inn \z_2)\tr(\veps_3\inn D)+16(\z_1  \inn
\z_2)k_1 \inn \veps_3 \inn k_2 \right.\cr &&\left.-16(p_3  \inn
\z_1)\z_2 \inn
 \veps_3 \inn k_2 -16(p_3  \inn \z_2)\z_1  \inn \veps_3
\inn k_1](\frac{\pi}{2})+\cdots \frac{}{}\right),\nonumber\eeqa
and $A(A_1,X_2,\tau_3)=0=A(A_1,X_2,h_3)$. The dots above
represents terms with more than two open string momenta, which
are related to the higher derivative terms in the effective
action. Now it is a simple exercise to check that above leading
terms of the decay amplitudes are reproduced exactly by the
corresponding terms in \reef{expand} as expected. To have overall
factors agreement as well, one should appropriately  normalize
fields in \reef{dbiac2}. This agreements insures us that the open
string tachyon potential in \reef{dbiac2} dose not have a linear
terms. Having such a term , the field theory action would produce
for example an amplitude for decaying two massless scalars to one
graviton, \ie $h^a{}_a\z_1\inn\z_2s/(1-2s)\sim
h^a{}_a\z_1\inn\z_2s$ which must be subtracted from string theory
amplitude $A(X_1,X_2,h_3)$. Consequently the result would not be
consistent with ${\cal L}(X,X,h)$ in \reef{expand}. Similarly,
above agreement predicts that the field theory does not have
coupling $TX^iX_i$. It is important to note that  the string
field theory in its present field variables has this coupling
\reef{vaat}, however using field redefinition \cite{AAT}, in
another field variables this coupling can be disappeared
\cite{vak}.

The decay amplitudes in section 3.2 have simple pole at
$s\rightarrow 0$, hence one may expect that field theory should
reproduce them. However,  as above analysis predicts the
tachyonic DBI action \reef{dbiac2} does not have couplings
$T\prt_aX^i\prt^a X_i$ or $TX^iX_i$,  hence the field theory
\reef{dbiac2} does not produce the simple massless poles of the
string theory amplitude. We expect that these massless  simple
poles are reproduced by the coupling $X^i\prt_i\tau$ or
$X^i\prt_i h^a{}_a$ which are in \reef{dbiac2} and  higher
derivative couplings like $\prt^a\prt_aT\prt_bX^i\prt^b X_i$ (or
$T\prt_a\prt_b X^i\prt^a\prt^b X_i$)  which are not included in
the tachyonic DBI action\footnote{A similar argument is presented
in \cite{AAT1} for consistency between sigma model approach
effective action and S-matrix elements.}. This means that the
on-shell amplitudes in section 3.2 should have a factor like
$1=(2k_1^2)^2$ that in off-shell case $2k_1^2\neq 1$ produces
higher derivative couplings in the field theory. Similarly the
contact terms of the amplitudes around $s\rightarrow 0$ also
describe the higher derivative terms which lead to these on-shell
contact terms.

Finally, the decay amplitudes in section 3.1 are smooth around
the top of tachyon potential $s\rightarrow 0$. Their expansion
around this point are
 \beqa
 A(T_1,T_2,\tau_3)&=&\left(\frac{ig_og_c}{12}\right)
 \left(\frac{}{}(-\pi s)+\cdots\frac{}{}\right),\nonumber\\
A(T_1,T_2,h_3)&=&\left(\frac{ig_og_c}{12}\right)
\left(\frac{}{}\tr(\veps_3
\inn D)(-\pi s)+16(k_1 \inn \veps_3 \inn
k_2)(\frac{\pi}{2})+\cdots \frac{}{}\right),\nonumber \\
A(T_1,T_2,\Phi_3)&=&\left(\frac{ig_og_c}{12\sqrt{24}}\right)
\left(\frac{}{}\tr(D)
(-\pi
s)-4(2-s)(\frac{\pi}{2})+\cdots\frac{}{}\right),\nonumber\eeqa
where dots represent terms that we expect them to be related to
higher derivative terms. Using the fact that in this case
$s=1+2k_1\inn k_2$, one can easily verifies that these contact
terms are reproduced exactly by the following couplings extracted
from \reef{dbiac2}: \beqa {\cal L}(T,T,\tau)&=&-\pi
T_p\,\tau\left(-T^2+\alpha'\prt_a
T\prt^a T\right),\nonumber\\
{\cal L}(T,T,h)&=&-\pi T_p\left(-\frac{1}{2}h^a{}_a
T^2+\frac{\alpha'}{2}h^a{}_a\prt_b T\prt^b T-\alpha'h^{ab}\prt_a
T\prt_b T\right),\nonumber\\
{\cal L}(T,T,\Phi)&=&-\pi
T_p\Phi\left(-\frac{(p-11)}{12}T^2+\frac{(p-13)}{12}\alpha'\prt_a
T\prt^a T\frac{}{}\right),\nonumber \eeqa where in writing the
last line we have used the fact that vertex operators in the
string amplitudes correspond to Einstein frame metric, whereas,
the metric in the tachyonic DBI action is string frame metric.
The relation between them is $g_S=e^{\phi/6}g_E$. The couplings in
the first line above indicates that the tachyon has mass
$m^2=-1/\alpha'$, as expected. The couplings in the second line
indicates that the graviton couples to tachyon by making
covariant the free action of tachyon. However, the couplings in
the last line  have a unique signature that imply the kinetic
term of tachyon  should appear  under square root of a
determinant as in \reef{dbiac2}. This ends our illustration of
the complete consistency between S-matrix elements at
$s\rightarrow 0$ limit and tachyonic DBI action.
\section{Discussion}
In this paper, using gauge invariant operators corresponding to
on-shell closed string states, 3-point interaction of open string
field theory and standard Feynman rules, we have explicitly
evaluated the amplitude for decaying two on-shell open string
states in level zero or one and one closed string tachyon or
graviton. In our calculation, the off-shell states propagating
between the on-shell closed string state and two open string
states in level more than two are truncated. We have then
evaluated explicitly the same amplitudes in the bosonic string
theory. Using an expansion for the beta function in these
amplitudes, we have expand the amplitudes in terms of a tower  of
simple poles reflecting the off-shell states propagating between
one on-shell closed string  and two open string states, like in
string field theory case. Truncating poles with more than level
two, we have found that the results are exactly the same as in
string field theory side up to some  contact terms. To compare
the contact terms of the amplitude in two theories as well, one
needs the amplitude in string field theory side to have all
infinite number of massive poles. This because individual contact
terms of each massive pole has contribution to contact terms of
the whole amplitude. It would be interesting then to find an exact
tree level amplitude in string field theory side and compare with
the exact form of the amplitude in the bosonic string theory
found in section 3.  Similar comparison has been done for the
scattering amplitude of four open string tachyons in
\cite{giddings}. Lastly, we have expanded the beta functions in
the exact disk level decay amplitudes in the bosonic string theory
case around $s\rightarrow 0$ and found that their leading order
terms can be fully described by the tachyonic DBI action
\reef{dbiac2}.

The tachyonic DBI action produces only the leading terms of
S-matrix elements expanded  around top of the tachyon potential
(not $\alpha'\rightarrow 0$ limit). Hence one may expect that the
other terms of expansion have significant effect.  So the field
theory effective action of string theory at the top of the
tachyon potential should have those contact terms as well, \ie
the action should have higher derivative terms as well.
Interestingly, Sen has  shown that some exact result of string
theory, like production of  a pressure-less gas with non-zero
energy density at the late time of the tachyon
condensation \cite{sen1}, can be derived also from the tachyonic
DBI action \cite{mrg0,EAB} at minimum of the tachyon potential.
Hence one may conclude  that, as an effective action of string
theory, although the higher derivative terms have important
impact at the top of the tachyon potential, they have little
effects at the minimum of the potential. Therefore, the tachyonic
DBI action derived by studying S-matrix elements at the top of
potential may be, in fact, effective action of string theory at the
minimum of the tachyon potential.

In our calculation of decay amplitude we have ignored recoil of
D-brane which allows us to ignore momentum conservation in the
transverse directions and hence keep $k_1\inn k_2$ arbitrary. This
procedure of relaxing momentum conservation is in fact a
particular way of off-shell extension of the S-matrix elements.
It would be interesting also to extend the on-shell conditions
$k_1^2=1/2=k_2^2$ to off-shell values as well. This might be
possible by comparing the S-matrix elements in bosoinc string
theory with the corresponding off-shell amplitudes in the string
field theory side. In this way one may be able to find higher
derivative corrections to the  action \reef{dbiac2}.
Alternatively, one may start from beginning with an off-shell
approach to effective action like sigma model approach \cite{AAT1}.
In this approach one directly reads effective action from a sigma
model partition function. It would be interesting then to see if
the tachyonic DBI action \reef{dbiac2} and its higher derivative
corrections can be extracted in this way. We hope to answer
these questions in our future works.

{\bf Acknowledgments}: We would like to acknowledge useful
conversations with M. Alishahiha. G.R.Maktabdaran was supported by Birjand
university.

\end{document}